\documentclass[prl,twocolumn,showpacs,preprintnumbers]{revtex4}
\usepackage{graphicx,psfrag,citesort,dcolumn,bm,amsmath,amssymb}
\usepackage[dvips]{color}

\usepackage{textcomp} 
\newcommand{\Notiz}[1]{\ifdim\overfullrule>0pt
  \marginpar[{\raggedleft\textcolor[rgb]{1,0,0}{#1}}] 
  {\raggedright\textcolor[rgb]{1,0,0}{#1}}{}\fi} 

\begin{document}

\title{Mechanics of bundled semiflexible polymer networks}

\author{O. Lieleg$^1$, M. M. A. E. Claessens$^1$, C. Heussinger$^2$,
  E. Frey$^2$ and A. R.  Bausch$^1$} \affiliation{$^1$Lehrstuhl f\"ur
  Biophysik E22, Technische Universit\"at M\"unchen,
  James-Franck-Stra\ss e 1, 85748 Garching, Germany, $^2$Arnold
  Sommerfeld Center (ASC) for Theoretical Physics and Center for NanoScience (CeNS),
  Ludwig-Maximilians-Universit\"at M\"unchen, Theresienstra\ss e 37,
  D-80333 M\"unchen, Germany}

\date{\today}

\begin{abstract}
  While actin bundles are used by living cells for structural
  fortification, the microscopic origin of the elasticity of bundled
  networks is not understood. Here, we show that above a critical
  concentration of the actin binding protein fascin, a solution of
  actin filaments organizes into a pure network of bundles. While the
  elasticity of weakly crosslinked networks is dominated by the affine
  deformation of tubes, the network of bundles can be fully understood
  in terms of non-affine bending undulations.\end{abstract}


\maketitle

The mechanical properties and dynamic organization of the cytoskeleton determine
the morphology and mechanical response of eukaryotic cells.  To ensure
adaptability of both organization and mechanics cells exploit the dynamic
interplay between semi-flexible polymers such as microtubules and actin
filaments using a multitude of associated binding proteins. In particular, the
local elastic properties are regulated by the activation of auxiliary proteins
which e.g. cross-link and/or bundle the filamentous networks into complex
scaffolds. Given the importance of the actin cytoskeleton for force generation
and transduction there is much interest in understanding the mechanical
properties of different network structures and the physical origin of the
transitions between them. This is best studied in \textit{in vitro} model
systems~\cite{Bausch2006}. In the absence of cross-links actin solutions are
successfully described by the spatial confinement of thermal bending undulations
upon affine tube deformation~\cite{Hinner1998}. Cross-linked semiflexible
polymer networks, on the other hand, are in general dominated by an interplay
between polymer stretching and bending modes, the precise form of which, as well
as the degree of non-affinity, strongly depends on the network
microstructure~\cite{Heussinger2006}.
So far the mechanical response of highly cross-linked actin networks, also in
the presence of bundles and composite phases, has mainly been described assuming
purely affine entropic stretching deformations~\cite{Gardel2004, MacKintosh1995,
  Shin2004, Tharmann2006a}. 
However, an applied tension can stretch a thermally undulating polymer only as
far as there is excess contour length available. As the maximal amount of stored
length is inversely proportional to the persistence length, entropic stretching
is suppressed in networks of stiff polymer bundles, where the persistence length
grows with bundle size~\cite{Claessens2006,Bathe2006}.  Moreover, the highly
non-linear nature of the force-extension relation of semi-flexible polymers
implies that linear elasticity is applicable as long as only a fraction of the
total excess length is pulled out. As an alternative the recently introduced concept
of the ''floppy modes'' may be better suited to describe the polymer elasticity in 
situations where entropic effects are suppressed~\cite{Heussinger2006a}. 
These floppy modes constitute bending excitations which, unlike the affine 
stretching deformations, retain a highly non-affine character.

In this Letter we show that above a critical concentration of the actin binding
protein (ABP) fascin, a solution of actin filaments organizes into a homogeneous
network whose building blocks are bundles only. At low cross-linker
concentration, the network response is dominated by the affine deformation of reptation
tubes and the ensuing changes in confinement free energy~\cite{Frey2001}. The observed
mechanical and structural transition between both phases can be described by a
simple relation between the ABP concentration and the entanglement length. It is
proposed to rationalize the scaling of the elastic modulus in the bundled regime
in terms of the floppy mode picture. A model based on affine stretching deformations 
only fits the data if additional assumptions about the bundle structure are made.

G-actin is obtained from rabbit skeletal muscle and stored in
lyophilized form at -21~$^\circ$C~\cite{Spudich1971}. For measurements
the lyophilized actin is dissolved in deionized water and dialyzed
against G-Buffer (2~mM Tris, 0.2~mM adenosine triphosphate (ATP), 0.2~mM CaCl$_2$, 0.2~mM dithiothreitol (DTT)
and 0.005~\% NaN$_3$, pH~8) at 4~$^\circ$C. The G-actin solutions are
kept at 4~$^\circ$C and used within seven days of preparation. The
average length of the actin filaments is controlled to 21~\textmu m
using gelsolin which is prepared from bovine plasma serum following~\cite{Sakurai1990}. 
Recombinant human fascin (55~kD) was prepared by a modification of the method
of~\cite{Ono1997} as described by~\cite{Vignjevic2002}. In the
experiments the molar ratio $R$ between fascin and actin, $R =
c_f/c_a$, is varied over almost three decades.

To resolve the structure and mechanical properties of actin/fascin-networks actin is polymerized
in F-buffer (2~mM Tris, 2~mM MgCl$_2$, 0.2~mM CaCl$_2$, 0.2~mM DTT,
100~mM KCl and 0.5~mM ATP, pH 7.5). For fluorescence microscopy filaments are stabilized with
tetramethyl rhodamine iso-thiocyanate (TRITC) phalloidin; either labeled reporter filaments (1 per 400) or
continuous labeling is used at distinct amounts of fascin. To avoid
photobleaching 0.6~\textmu M Glucoseoxidase, 0.03~\textmu M Catalase
and 0.01~M Glucose are added. The samples for transmission electron
microscopy (Philips EM 400T) are adsorbed to glow-discharged
carbon-coated formvar films on copper grids. The samples are washed in a
drop of distilled water and negatively stained with 0.8~\% uranyl
acetate; excess liquid is drained with filter paper. The viscoelastic
response of actin/fascin-networks is determined by measuring the
frequency-dependent viscoelastic moduli $G'(\omega)$ and $G''(\omega)$
with a stress-controlled rheometer (Physica MCR 301, Anton Paar, Graz,
Austria) within a frequency range of three decades. Approximately
520~\textmu l sample volume are loaded within 1~min into the rheometer
using a 50~mm plate-plate geometry with 160~\textmu m plate
separation. To ensure linear response small torques are applied. Actin
polymerization is carried out in situ, measurements are taken 60~min
after the polymerization was initiated.

\begin{figure}[htbp]
\includegraphics[width = \columnwidth]{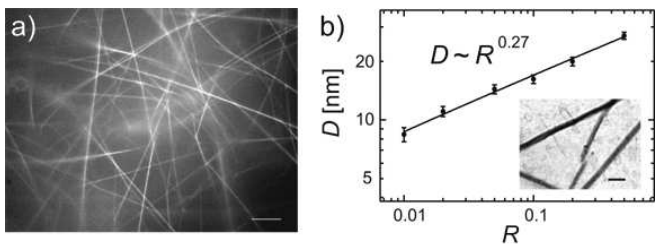}
  \caption {a) Fluorescence micrograph of an actin/fascin network 
    (0.1 mg/ml actin): for high fascin concentrations a purely bundled network is formed
    (scale bar is 10~\textmu m). \protect \\ b) From TEM pictures (inset, scale
    bar is 0.2~\textmu m) a scaling relation for the average bundle
    diameter $D$ is obtained.}
        \label{Fig1}
\end{figure}

\begin{figure}
\includegraphics[width= \columnwidth]{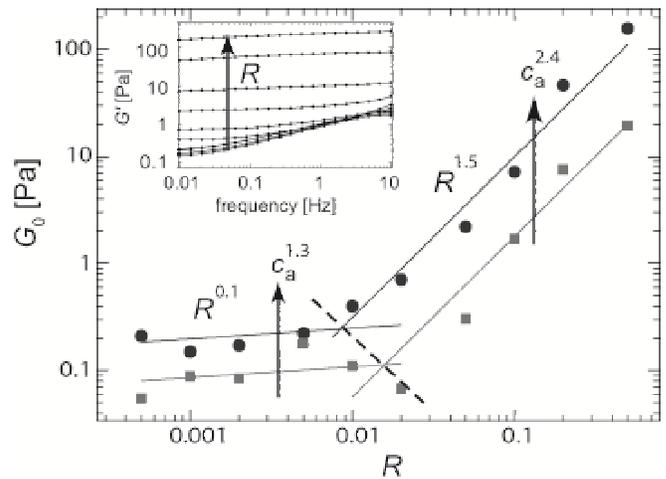}
  \caption {Plateau modulus $G_0$ as a function of the molar ratio 
    $R$ of fascin with respect to actin for two different concentrations of actin: 0.4~mg/ml (circles) and 0.2~mg/ml
    (squares). The dependence of $G_0$ on $c_a$ is obtained by scaling the fits for the 0.2~mg/ml actin data
    upon the 0.4~mg/ml data points. The dashed line shows the boundary separating the two scaling regimes. 
    The original frequency spectra for 0.4~mg/ml actin at different cross-linker concentrations
    ($R$~=~0, 0.001, 0.002, 0.005, 0.01, 0.02, 0.05, 0.1, 0.2, 0.5) are depicted in the inset.}
        \label{Fig2}
\end{figure}

Fluorescence images show that in the presence of high concentrations
of fascin, actin filaments organize into a network of bundles (Fig.~\ref{Fig1}a) 
while below a critical value $R^* \approx 0.01$ no
bundles can be observed. Both fluorescence and transmission electron
microscopy do not show any sign of composite phases or microdomains as
observed in the presence of other ABPs~\cite{Tempel1996, Shin2004,
  Wagner2006}. Moreover, the existence of a purely bundled phase is demonstrated 
by a cosedimentation assay~\cite{supplement}. 
The bundles formed are very long ($>$ 100~\textmu m)
and straight which is consistent with the measured bending rigidity
$\kappa$~\cite{Claessens2006}. TEM micrographs reveal that above $R^*$
the actin/fascin bundle thickness $D$ and therefore the number of
actin filaments per bundle, $N$, increases weakly with $R$
(Fig.~\ref{Fig1}b).  The bundle thicknesses are extracted from the
TEM-micrographs by fitting a Gaussian to the intensity profiles,
obtaining a scaling of $D~\sim~N^{1/2} \sim~R^{x}$ with
$x~=~0.27$.
Concomitant with the structural changes the viscoelastic properties of
the network alter: with increasing $R$ both the storage modulus
$G'(\omega)$ and the loss modulus $G''(\omega)$ increase over the
whole frequency range probed. The storage modulus $G'(\omega)$
exhibits a plateau at low frequencies, while the loss modulus
$G''(\omega)$ reveals a well-defined minimum which shifts to higher
frequencies with increasing $R$. The plateau modulus $G'$(10 mHz)
plotted against $R$ shows two distinct regimes in the elastic
response. At low $R$, $G_0$ is only slightly dependent on $R$, $G_0
\sim R^{0.1~\pm~0.1}$, while above a critical value $R^{**}$, $G_0$ increases
with $G_0 \sim R^{1.5~\pm~0.2}$ (Fig.~\ref{Fig2}). This exponent fits the 
data for both actin concentrations probed ($c_a$~=~0.2~ mg/ml and
$c_a$~=~0.4~mg/ml). The transition point $R^{**}$ agrees well with
the structural transition at $R^*$ observed in microscopy. Below $R^*
= R^{**}$ the plateau modulus scales with the actin concentration as
$G_0 \sim c^{1.3}_a$ suggesting that entanglements dominate the
elastic response~\cite{Hinner1998}. Above $R^*$ a different scaling
regime occurs with $G_0 \sim c_a^{2.4}$.

With the observed scaling behavior $G_0(R, c_a)$ the plateau modulus
is parameterized in both regimes, before and after the structural
transition. At the cross-over concentration $R=R^*$ these two
parameterizations have to be equal. This uniquely determines the
scaling of $R^*$ with the actin concentration, $R^*\sim
c_a^{-0.79}$, which results in the constraint $c_f \cdot
c_a^{-0.21} \sim 1$. This can be approximated to $c_f \cdot l_e^{1/2} \sim 1$
using the entanglement length $l_e \sim c_a^{-2/5}$.  This
surprisingly simple criterion for the bundling transition defies an
obvious explanation and a detailed theoretical model is still lacking.
It would need to account for the subtle interplay between confinement
free energy of polymers in both the bundle and the network as well as
the binding enthalpy of the cross-linking proteins.

The mechanical properties inside the bundled regime may, on the other
hand, be understood in terms of the non-affine floppy mode
model~\cite{Heussinger2006a}, where network elasticity is attributed
to bending modes of wavelength comparable to the distance between
cross-links, $l_c$, and with stiffness $k_\perp \sim \kappa / l_c^3$.
In this picture typical deformations of the network do not follow the
macroscopic strain affinely but scale as $\delta_\text{na} \sim \gamma
L_B$, where $L_B$ is a constant length over which an individual bundle
within the network can be assumed to be straight.  From our
fluorescence and TEM pictures, we would expect this length to be
comparable to the bundle length. As a consequence the linear elastic
modulus reads
\begin{eqnarray}
  G_0 \sim \nu k_\perp \delta_\text{na}^2 
\end{eqnarray}
with the polymer density $\nu \sim 1/\xi^2 l_c$. This model can be tested 
by relating the structural parameters of the network, mesh size $\xi$ 
and $l_c$, and the bending elasticity $\kappa$ of the bundle segment to 
the concentration of actin and fascin monomers ($c_a, c_f$).

The structural information obtained by TEM and fluorescence microscopy justifies
the assumption that the bundles form an isotropic network similar to an
entangled structure of single filaments. With increasing $R$, filaments and
smaller bundles reorganize to form larger bundles that are spaced further apart.
The mesh size $\xi$ of this self-similar structure therefore depends
on $R$ as $\xi \sim \xi_0 N^{1/2}$, where $\xi_0 \sim c_a^{-1/2}$ is the
mesh size of the filamentous network. Cross-linking will typically occur on the
scale of the entanglement length $l_e$, which plays the role of a distance
between bundle-bundle intersections (entanglement points). Since on average only
a fraction of those will be occupied we can assume that distances between
cross-links along the same bundle are given by $l_c~\sim~R^{-y}
l_e$~\cite{Tharmann2006a, Shin2004}. Doubling the cross-linker concentration $R$
should halve the distance between them, suggesting an exponent $y \approx 1$.

For a description of the elastic properties of the bundles it is necessary to
realize that fascin only leads to loosely coupled bundles, where bending is
dominated by the shear stiffness of the cross-linking
proteins~\cite{Claessens2006, Bathe2006}. The key quantity in this context is
the bundle coupling parameter $\alpha(l_c) = (l_c/b)^2$, where the length-scale
$b\sim\delta^{1/2}$ encodes the properties of the ABP's inside the bundle, in
particular via the average distance $\delta$ between cross-links. In general,
$\delta$ will depend on the concentrations $c_f$ and $c_a$, however, the precise
relationship is not known. From fluorescence images the mesh size of the bundled 
network at $R = 0.5$ can be approximated which allows to calculate $l_e$ and thus
$l_c$. From this one can estimate the coupling parameter to be $\alpha~>~1$ 
for the whole bundle regime, implying that the effective bundle bending
stiffness $\kappa$ acquires a wavelength dependence~\cite{Bathe2006}, leading to
$\kappa(\lambda) \sim N \kappa_f \alpha(\lambda)$, where $\lambda$ is
the wavelength of the deformation mode. This stands in marked contrast to what
is known for single filaments or scruin-bound bundles where a fully coupled
bending regime, $ \kappa\sim N^2 \kappa_f$, has been assumed~\cite{Shin2004a}.
The wavelength dependence of the bundle stiffness has far reaching consequences
for the static as well as dynamic properties of semi-flexible polymer
bundles~\cite{Heussinger}.  In particular, it implies that the entanglement
length $l_e$ has to be reevaluated. As it derives from the suppression of long
wavelength fluctuations by confining a polymer into a tube~\cite{Odijk1983} it
is highly sensitive to a wavelength dependent $\kappa(\lambda)$. This results
in $l_e^3 \sim (Nl_p)\xi^4/b^2$, which is different from the usual expression
$l_e^5 \sim l_p\xi^4$ valid for single filaments, where in both cases $l_p$ = 17~\textmu m~\cite{LeGoff2002} 
denotes the persistence length of a single actin filament. Combining the above results
and setting the deformation mode length $\lambda$ equal to $l_c$ one finally
arrives at the following prediction
\begin{eqnarray}
  G_0 \sim R^z c_a^w\cdot\delta^{-1/3},
\end{eqnarray}
where the exponents are given by $z=2y-4x$ and $w=7/3$. Thus the scaling
exponent of the plateau modulus can be related to parameters describing the
microstructure such as the scaling of the mesh size as well as the dependence of
the bundle thickness and elasticity on $R$. From our measurements of $x~=~0.27$
and $z~=~1.5$, and by assuming $\delta$ to be a constant, a value of $y~=~1.29$
is obtained, which is in reasonable agreement with the expected
$y~\approx~1$. This result is largely insensitive to the assumption of
constant $\delta$, since by assuming $\delta$ to change according to simple Langmuir
kinetics an exponent $y~\approx~1.21$ is obtained.

\begin{figure}
\includegraphics[width= \columnwidth]{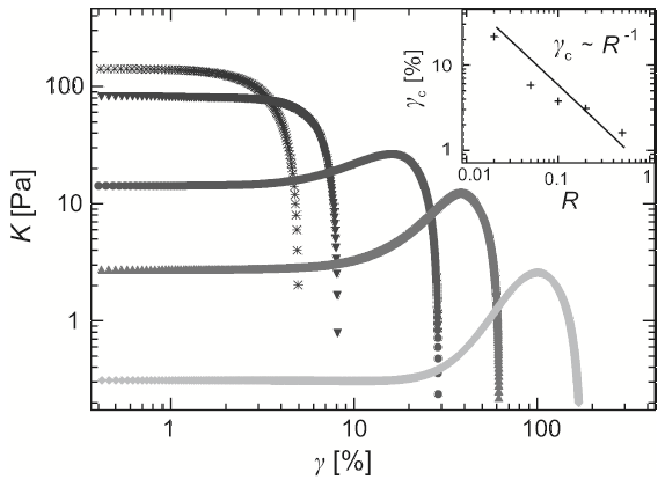}
  \caption {Differential modulus $K=d\sigma/d\gamma$ plotted versus deformation $\gamma$ 
    for fascin networks in the bundle phase ($c_a$ = 0.4 mg/ml and increasing $R$: 
    diamonds $R~=~0.02$, upright triangles: $R~=~0.05$, circles: $R~=~0.1$, downright triangles: 
    $R~=~0.2$, stars: $R~=~0.5$). The inset shows the critical strain $\gamma_c$ in dependence on $R$.}
  \label{Fig3}
\end{figure}

To further characterize the elastic properties in the bundled regime, the
non-linear elasticity of the network is investigated. For samples with $R~>~R^*$
a constant shear rate is applied and the resulting stress is reported. From
the smoothed $\sigma(\gamma)$ relation the numerical derivative yielding the
differential modulus $K = \partial\sigma/\partial\gamma$ is calculated
(Fig.~\ref{Fig3}).  For small strains of $\gamma=1~-~10~\%$ linear behavior is
observed, where the differential modulus follows $K~\sim~R^{1.5}$ in agreement
with our oscillatory measurements. A non-linear response is observed above
$\gamma_c$, which is determined as the strain at which $K$ deviates by 5~\% from
its value in the linear regime. Up to $R~=~0.1$ a strain hardening occurs while
for very high values of $R$ the linear regime is directly followed by strain
weakening. The disappearance of the strain hardening at high concentrations of
fascin might be the result of the rupturing of fascin-actin bonds - very similar
to what was reported for rigor-HMM-networks~\cite{Tharmann2006a}.
The floppy mode description also has implications on the onset of the non-linear
behavior. As has been argued in~\cite{Heussinger2006a} large strains necessarily
lead to stretching even if the deformations were only of bending character. The
bundle stretching $\Delta$ is related to the transverse displacement
$\delta_{\rm na}$ by simple geometric considerations as $l_c^2+\delta_{\rm na}^2
= (l_c + \Delta)^2$. The floppy mode description thus applies as long as this
stretching is small compared to the available thermal excess length $\Delta
\Lambda \sim l_cb/Nl_p$~\cite{Heussinger}.  This defines a critical strain
$\gamma_c \sim l_c(b/N)^{1/2} \sim R^{-y+x}b^{-1/6} \sim
R^{-1.0}\delta^{-1/12}$ for the onset of non-linear effects. The scaling with
$R$ is in excellent agreement with our measurements (see inset of
Fig.~\ref{Fig3}). The weak dependence on the cross-linker spacing
$\gamma_c\sim\delta^{-1/12}$ also implies that this result is insensitive
towards any putative dependence on the fascin concentration via
$\delta=\delta(c_f)$.

On the other hand, if one were to apply an affine stretching model, a different
picture emerges, where $\delta(c_f)$ has to be tuned to obtain a reasonable data
fit.  In such a model one would assume the modulus to be given by $G_{\rm aff} \sim
\nu k_{\parallel} \delta_{\rm aff}$, where $k_{\parallel}\sim
\alpha^{3/2}(N\kappa_f)^2/l_c^4$ is the stretching stiffness of the
bundle~\cite{Heussinger}. The deformations are assumed to be affine, implying
$\delta_{\rm aff}\sim \gamma l_c$.  The modulus thus reads as $G_{\rm aff} \sim
R^{2x}c_a\cdot \delta^{-3/2}$~\cite{supplement} while the critical strain $\gamma_c\sim
\delta^{1/2}R^{-2x}$ is obtained by equating $\Delta\Lambda$ with the affine
deformation $\delta_{\rm aff} \sim \gamma l_c$.  This model can only fit the data by assuming
$\delta \sim c_f^{-\beta}$ with an exponent in the range of $\beta = 0.6~-~0.9$, such
that for $\beta = 0.6$ the $R$-dependence of the modulus and for $\beta = 0.9$ the 
$c_a$-dependence of the modulus is reproduced. To finally decide whether or not the 
application of an affine stretching model is equally successful as the floppy mode 
approach, $\delta(c_f)$ would have to be determined by scattering experiments.

In summary, on the basis of a combined microscopy and rheology study we have
shown that the actin binding protein fascin mediates a transition from an
entangled polymer solution to a homogeneously cross-linked bundle phase. These
phases differ both in structure and mechanical properties. The location of the
transition is given by a simple relation between the ABP concentration and the
entanglement length. Moreover this transition point seems to be more general since 
it also occurs at similar ABP concentrations in other systems such as isotropically 
cross-linked networks or even composite networks~\cite{Wagner2006, Tharmann2006a, Gardel2004}. 
The transition is a consequence of the interplay between the chemical kinetics
of the binding proteins, the bending rigidity of the polymers and the entropic
forces between those components. How the concerted action of those driving
forces leads to such a structural transition is an interesting theoretical
problem. 
The elasticity in the bundled phase is well explained in terms of a
recently developed floppy mode picture~\cite{Heussinger2006a}. We have
argued that in the absence of a significant amount of stored length in
the bundles, non-affine bending is the dominant low energy excitation.
It explains both the linear elasticity and the onset of non-linear
behavior. This model has to be seen as an alternative to affine models
where the elastic response is due to pulling out stored length
fluctuations. While the elasticity of isotropic networks may be
predominantly determined by such an entropic stretching of single
filaments between the cross-linking points, we suppose that the
elastic response of composite phases may instead be dominated by
non-affine deformations of bundles as described by floppy modes.  The
detailed understanding of the presented purely bundled network,
composed of shear dominated bundles, provides a benchmark for
addressing the further challenge to describe the mechanics of
networks, which are dominated by structural heterogeneities.

We thank M. Rusp for the actin preparation.  This work was supported by Deutsche Forschungsgemeinschaft through the DFG-Cluster of Excellence Nanosystems Initiative Munich (NIM) and SFB 413. Oliver Lieleg was supported by CompInt in the 
framework of the ENB Bayern.

\end{document}